\documentclass[aps,epsfig,pre,twocolumn,nofootinbib]{revtex4}
\usepackage[dvips]{graphicx}
\usepackage[dvips]{color}
\usepackage{subfigure}
\usepackage{epsfig}
\usepackage{dcolumn} 
\usepackage{bm} 
\usepackage{amsmath}
\usepackage{amssymb}
\newcommand\beq{\begin{equation}}
\newcommand\bear{\begin{eqnarray}}
\newcommand\eeq{\end{equation}}
\newcommand\eear{\end{eqnarray}}

\begin{document}

\title{Spatial modulation of the composition of a binary liquid  near a repulsive wall}
\author{Shibu Saw$^{1}$\footnote{shibu@physics.iisc.ernet.in}, S. M. Kamil$^{1,2}$\footnote{kamil.syed@snu.edu.in}, and Chandan Dasgupta$^{1}$\footnote{cdgupta@physics.iisc.ernet.in}}
\affiliation{$^{1}$Department of Physics, Centre for Condensed Matter Theory, Bangalore 560012, India\\ 
$^{2}$Department of Physics, School of Natural Sciences, Shiv Nadar University, Gautam Budh Nagar 201314, India }

\date{\today}

\begin{abstract}{
When a binary liquid is confined by a strongly repulsive wall, the local 
density is depleted near the wall and an interface similar to that between 
the liquid and its vapor is formed. This analogy suggests that the composition 
of the binary liquid near this interface should exhibit spatial modulation 
similar to that near a liquid-vapor interface even if the interactions of 
the wall with the two components of the liquid are the same. The Guggenheim 
adsorption relation quantifies the concentrations of two components of a 
binary mixture near a liquid-vapor interface and qualitatively states that 
the majority (minority) component enriches the interface for negative (positive) 
mixing energy if the surface tensions of the two components are not very
different. From molecular dynamics simulations of binary mixtures with 
different compositions and interactions, we find that the Guggenheim relation 
is qualitatively satisfied at wall-induced interfaces for systems with negative 
mixing energy at all state points considered. For systems with positive mixing 
energy, this relation is found to be qualitatively valid at low densities, 
while it is violated  at state points with high density where correlations 
in the liquid are strong. This observation is validated by a calculation of 
the density profiles of the two components of the mixture using density 
functional theory with the Ramakrishnan-Yussouff free-energy functional. 
Possible reasons for the violation of the Guggenheim relation are discussed. 
} 
\end{abstract}

\pacs{xx}

\maketitle

\section{Introduction}

Equilibrium and dynamic properties of strongly confined liquids are known~\cite{confinement} to be substantially different from
those of their bulk counterparts. Confinement of a single-component liquid by a repulsive wall produces a modulation of the 
local density (layering) of the liquid near
the wall. This phenomenon has been studied extensively in experiments~\cite{expt}, simulations~\cite{simul} and theoretical
analyses~\cite{theory}. The effects of confinement on binary mixtures have received less attention in the existing literature.
When a binary liquid is confined by a wall that has different interactions with the two components of the liquid, the
presence of the wall causes the concentrations of the components near the wall to be different from those in the bulk. This can lead
to surface-induced phase separation and related phenomena~\cite{puri}. Even if the interactions of the wall with the two
components are the same, the presence of the wall may lead to a modulation of the local composition in addition to the more familiar
modulation of the total density. This possibility is suggested by the following argument. The presence of a sufficiently 
repulsive wall leads to the formation of a region near the wall where the density of the liquid is considerably smaller than 
that in the bulk and the situation would be similar to that of a liquid in contact with its vapor. It is known from earlier 
work~\cite{GibbsAdsorption,Guggenheim1944} that the composition of a multicomponent
liquid near a liquid-vapor interface generally differs from that in the bulk. So, the local composition of a binary liquid near a 
strongly repulsive wall that does not distinguish between the two components is expected to exhibit spatial variation. In this paper,
we study this spatial variation using computer simulations and density functional theory.

There are many physical systems in which modulations of the density and composition of a multicomponent liquid near a 
repulsive wall play an important role. Binary mixtures are often used in numerical studies of 
glass-forming  liquids 
because it is difficult to supercool single-component model liquids with simple interactions. There exist many simulations~\cite{tanakaNmat2011} of 
dense binary liquids confined by repulsive walls. The modulation of the composition of the mixture near the confining walls plays an
important role in the behavior of these systems. This aspect is particularly important in understanding the properties of
vapor-deposited glasses which, depending on the substrate temperature and deposition rate,  can 
have high density, low enthalpy and higher mechanical moduli than liquid-cooled 
glasses~\cite{Ediger2012JPCLett,Swallen2007Science,Ediger2007JCP,Ramos2011JPCB,Ediger2008JPCB,dePablo2011,dePablo2013,Ediger2010AdvMater}.
The anisotropy and higher density of vapor-deposited glasses make them potentially important for industrial applications such as optoelectronics~\cite{Ediger2012JPCLett,Swallen2007Science,Yokoyama2008APL}. 
These glasses typically have two or more components. As discussed above, the local concentrations
of different components near the substrate, as well as near the interface with the vapor, 
are, in general, not the same as those in the bulk. This change of 
composition should strongly affect the characteristics of vapor-deposited glasses.
Such glasses may have properties that are quite different from those of liquid-cooled 
glasses that have the same composition throughout the sample~\cite{dePablo2013condmat,Debenedetti2011JCP}.
It is therefore important to understand the compositional behavior 
of  a mixture near surfaces in studies of vapor-deposited glasses. 
Our study of the compositional behavior of a binary liquid
near a repulsive wall is a step in this direction.

Another context in which the questions addressed in our study are relevant is in determining whether a binary mixture would be a 
good metallic glass former. The mixing energy is one of the factors that decide the answer to this question.
The mixing energy of an $A + B$ mixture is defined as $\omega'=z\omega$, where
$z$ is the coordination number and $\omega = [E_{AB} - (E_{AA} + E_{BB})/2]$, with
$E_{AA}$, $E_{AB}$ and $E_{BB}$ the minimum energies between
the $A$-$A$, $A$-$B$ and $B$-$B$ pairs, respectively. Alloys with negative (positive) mixing energy are believed to be good (bad)  
glass formers~\cite{Ma2005ProgMatterSci}. 
The relation between the mixing energy and glass-forming ability of a metallic binary mixture
in the presence of surfaces has been studied recently~\cite{Harrowell2013NatMat}. 
Here we study the compositional behavior of binary mixtures with both negative 
and positive mixing energies.

As noted above, the behavior of a binary liquid near a strongly repulsive wall is 
similar to that near a liquid-vapor interface. There exist several experimental 
studies~\cite{Shpyrko2005,Shpyrko2004,Tostmann2000PRL,Lei1996JCP,Regan1997PRB,EDiMasi2001PRL} of
the local density and composition of liquid metal alloys near liquid-vapor interfaces. These studies show
that liquid metal alloys exhibit density as well as 
composition modulation near an interface. The results of our study would be useful for understanding the behavior observed in these experiments.

Classical density functional theory (DFT)~\cite{dftrev} has been used~\cite{dft1,dft2} in the past
to study the density profiles of binary mixtures near walls that have different interactions with
the two components of the mixture.
We are not aware of any theoretical analysis of the spatial dependence of the composition of
a binary liquid near a repulsive wall that has the same interaction with the two components of the mixture. We therefore compare the results of our simulations with theoretical 
predictions for the similar problem of binary mixtures near a liquid-vapor interface.
Existing theories~\cite{GibbsAdsorption,Guggenheim1944} of the surface composition of binary liquids were formulated a long
time ago, but these theories continue to be used to understand the behavior found in
experiments~\cite{Shpyrko2005,Tostmann2000PRL,EDiMasi2001PRL} and simulations~\cite{Osborne}.  
The simplest of these theories is the
Gibbs adsorption rule~\cite{GibbsAdsorption},
which states that the component with the lower surface tension should accumulate at the
interface. This rule is valid for systems with zero mixing energy. The component with lower surface tension should enrich the interface also in the case where the 
surface tensions of the two components are very different and the mixing energy is small. 
When these two criteria (low mixing energy and very different surface tensions of the 
components) are not satisfied, the bulk concentrations of the two components 
play an important role in determining the surface concentrations.
For systems with non zero
mixing energy, surface concentrations 
of the two components are theoretically predicted from the Guggenheim 
relation~\cite{Guggenheim1944}, which states that the interface is enriched by the
majority (minority) components for negative (positive) mixing energy, provided 
the above-mentioned criteria are not fulfilled~\cite{Osborne}. Existing experimental results
for liquid metallic alloys~\cite{Shpyrko2005,Tostmann2000PRL,EDiMasi2001PRL} are 
qualitatively consistent with the predictions of the Guggenheim relation. We are not aware
of any other theoretical prediction for the liquid-vapor interfacial composition of binary
mixtures. 

In this paper, we examine in detail the spatial modulation of the composition of a
binary mixture near the interface created by the presence of a strongly repulsive wall that has
the same interaction with the two components of the mixture. 
We present the results of extensive simulations of the equilibrium behavior of several binary liquids confined by different kinds of repulsive walls. Our results provide detailed information
about how the spatial variation of the composition in the confinement direction depends on
physical parameters such as the structure of the repulsive wall, the interaction of the wall
with the particles in the liquid, the bulk composition of the liquid, the mixing energy, and 
thermodynamic parameters (density and temperature) of the bulk liquid.
We show that the spatial modulation of the concentrations of the 
two components near the wall-liquid interface is in qualitative
agreement with the prediction of the Guggenheim relation 
at all densities and temperatures considered if the mixing energy is negative. The
Guggenheim relation is also found to be qualitatively satisfied in binary liquids with 
positive mixing energy if the density is low.
However, we find a qualitative violation of the Guggenheim relation at high densities 
in two binary systems with positive mixing energy. A DFT calculation using
the Ramakrishnan-Yussouff free-energy functional~\cite{RY}, which takes into account the 
effects of short-range order in the liquid, is found to yield results that are
in agreement with those of our simulations for systems with both positive and negative
mixing energies. These results suggest that correlations present in the binary liquid at
high densities, which are manifested as pronounced short-range order, are responsible for the 
observed violation of the Guggenheim relation in dense binary liquids with positive
mixing energy.

The remainder of the paper is organized as follows. 
In Sec. II we describe the Guggenheim relation and present a simple derivation of
this relation in order to point out the approximations made in its derivation.
The systems studied and the simulation methods used in this work are described in Sec. III.
In Sec. IV we present the results obtained from our molecular 
dynamics (MD) and Monte Carlo (MC) simulations for several binary liquids confined
by different kinds of repulsive walls and compare these with the prediction of the Guggenheim relation.
In Sec. V we compare the density profiles near the interface obtained from 
classical DFT calculation and MD simulations.
Section VI contains a summary of the main results of our study
and a few concluding remarks.
\vspace{-6mm}\section{The Guggenheim Relation}
The Guggenheim relation~\cite{Guggenheim1944} that predicts the 
surface concentrations of the components of a regular mixture ({ i.e.} one with $\omega \ne 0$) 
has the form
\begin{eqnarray}
\gamma_A &+& \frac{k_BT}{a} \ln(\frac{x'}{x}) + \nonumber \\
&& \frac{\omega^{'}}{a}[l(1-x')^2-(l+k)(1-x)^2] \nonumber \\
 = \gamma_B &+& \frac{k_BT}{a} \ln(\frac{1-x'}{1-x}) + \nonumber \\
&& \frac{\omega^{'}}{a}[lx'^2-(l+k)x^2] \label{G2}
\end{eqnarray}
where $\gamma_A$ and $\gamma_B$  are the surface tensions of $A$ and
$B$ components, respectively, 
$a$ is the cross-sectional area of a particle (assumed to be the same for both kinds
of particles), and
$x$ and $x'$ are bulk and surface concentrations of $A$ component, respectively. 
Also, $lz$ and $kz$ 
are the coordination numbers in the lateral and normal planes to the 
interface ($l+2k=1$ for a single-layer interface),  $k_B$ is the 
Boltzmann constant, and $T$ is the absolute temperature.  For a 
face-centered-cubic (fcc) lattice, z = 12, l = 1/2 and k = 1/4;  these values 
are used in the present work. This relation is 
based on several assumptions such as the presence of a single-layer interface, 
only nearest-neighbor interactions, and similar sizes of the two components.

 For zero mixing energy ($\omega=0$), Eq.
 \eqref{G2} readily yields
\begin{equation}
x' = \frac{x}{1-(1-C)(1-x)}, \label{Gibbs1}\\
\end{equation}
where $C=\exp[a(\gamma_A-\gamma_B)/k_BT]$; $C \in (0,1)$ for 
$\gamma_A < \gamma_B$ while $C \in (1,\infty)$ for $\gamma_A > \gamma_B$ and 
values of $x$ and $1-x$ lie between 0 and 1 as they are bulk concentration 
of $A$ and $B$, respectively. From Eq. \eqref{Gibbs1}
it is evident that $x' > x$ for $\gamma_A < \gamma_B$ and $x' < x$ if 
$\gamma_A > \gamma_B$ and hence the component with lower surface tension 
dominates the interface. This is the { Gibbs adsorption law}.

\begin{figure}[t]
\includegraphics[scale=0.55,angle=0]{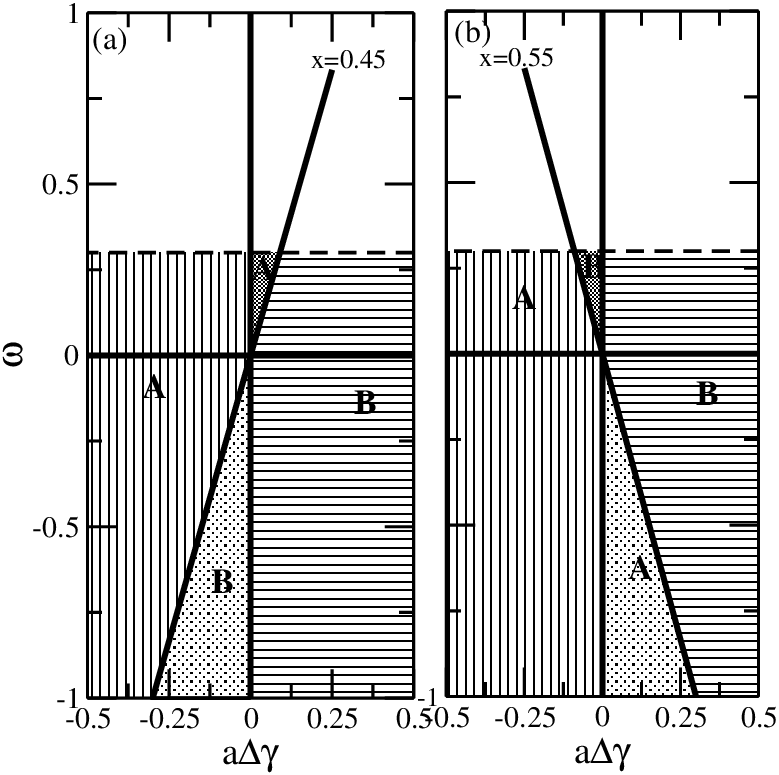}
\caption{Guggenheim relation is depicted in the $\omega$-$a\Delta \gamma$
parameter space. The straight lines passing through the origin show the parameter values at which
the surface concentration $x'$ and bulk concentration $x$ of component $A$ 
are equal for (a) $x=0.45$ and (b) $x=0.55$. In the region of parameter space to the right (left) 
of these lines, the surface layer is enriched by the $B$ ($A$) component. 
The enrichments in the shaded regions are contrary to the prediction of 
the Gibbs adsorption law. They also exhibit the dominance of the majority and minority components
for negative and positive mixing energies, respectively. The dashed lines schematically represent the boundary
above which bulk phase separation occurs.}
\label{Gugg_major}
\end{figure}

From Eq. \eqref{G2} one can obtain the loci of points in the $\omega$-$a\Delta\gamma$ 
parameter space ($\Delta\gamma=\gamma_A-\gamma_B$) with the same 
surface and bulk concentrations of component $A$ by setting 
$x'=x$. This is given by \cite{Osborne}
\begin{eqnarray}
\omega = \frac{a\Delta \gamma}{kz(1-2x)},
\end{eqnarray}
which is the equation of a straight line passing through the origin. 
The slope of this straight line 
increases as $x$ is increased from zero and changes sign as $x$ is increased beyond $0.50$. 
Figures. \ref{Gugg_major}(a) and \ref{Gugg_major}(b) exhibit lines of
equal concentration for  $x=0.45$ and $0.55$, respectively. In the region to the right of the line of 
equal concentration, the surface layer is enriched by $B$, while it is enriched by $A$
in the region on the left side of the line.
While $a\Delta\gamma$ tends to enrich the surface layer with the component with lower surface tension
($B$ for $a\Delta\gamma>0$ and $A$ for $a\Delta\gamma<0$), 
the mixing energy $\omega$ acts to reverse this trend in the region bounded by the
$\omega$-axis and the line of equal concentration (shaded regions 
in Fig. \ref{Gugg_major}). From Fig. \ref{Gugg_major} it is evident that for $\Delta \gamma \simeq 0$,
the majority (minority) component enriches the interface for negative 
(positive) $\omega$. The larger the value of $\omega$, the majority 
(minority) enrichment, contrary to the prediction of the Gibbs adsorption law 
occurs for larger $|a\Delta\gamma|$. Of 
course, for very large positive $\omega$, the system will phase separate 
into $A$-rich and $B$-rich regions in the bulk. This is schematically shown by the dashed lines in Fig. \ref{Gugg_major}.

We present here a simple derivation of the Guggenheim relation in order to bring out the assumptions and approximations
made in deriving Eq. \eqref{G2}. The binary mixture is described as an Ising model defined on a lattice.
Every site of the lattice is assumed to be occupied by a particle of type $A$ or $B$. The Ising variable $\sigma_i$ at lattice 
site $i$ takes the value +1 if the site is occupied by a particle of type $A$ and $-1$ if
it is occupied by a particle of type $B$. Defining $m_i$ as the thermal average of $\sigma_i$, the bulk concentration $x$ of $A$
particles is given by $x=(m+1)/2$, where $m$ is the value of $m_i$ for a bulk site. In molecular field theory~\cite{mft}, the
bulk magnetization $m$ satisfies the self-consistent equation
\begin{equation}
m=\tanh[\beta(Jzm+h)], \label{mft1}
\end{equation}
where $\beta \equiv 1/k_BT$, the interaction strength $J$ in the Ising model is given by $J=\omega/2$, and $h$ is a 
magnetic field required for fixing the bulk magnetization at the value corresponding to the prescribed value of $x$ (the value
of $h$ is related to the difference between the chemical potentials of $A$ and $B$ particles and to the difference between $E_{AA}$
and $E_{BB}$).
The self-consistent equation for the magnetization $m'=2x'-1$ of an Ising spin in the surface layer is
\begin{equation}
m'=\tanh\{\beta [Jz(lm'+km)+h']\}, \label{mft2}
\end{equation}
where the field $h'$ at the surface layer can be different from the field $h$ in the bulk. A few lines of algebra show
that these equations reduce to Eq. \eqref{G2} if the difference between the surface tensions 
$\gamma_A$ and $\gamma_B$ is identified with $2(h-h')/a$.

The following are the most important approximations made in the above derivation of the Guggenheim relation. (i) The total density is assumed to be
the same in the surface layer and in all bulk layers (each lattice site is assumed to be occupied by a particle of type $A$ or $B$), whereas the
presence of a repulsive wall is known to cause a strong modulation of the total density near the wall in strongly correlated liquids.
(ii) The composition is assumed to be different from the bulk value in only one surface layer. (iii) The derivation is based on molecular
field theory in which fluctuation effects are ignored. The issue of which of these approximations is likely to be responsible for a violation
of the Guggenheim relation in strongly correlated liquids will be discussed in Sec. VI.
 
\section{Simulation Details}
We have performed constant particle number, volume, and 
temperature  MD simulations for
$2400$ particles of binary Lennard-Jones (BLJ) mixtures consisting 
of $A$ and $B$ particles with different compositions at different temperatures and densities. 
In these simulations, the temperature was kept constant using the Brown-Clarke 
thermostat\cite{BCthermostat}. The system was confined in the $z$ direction
by two structureless non-preferential ({ i.e.} same for $A$ and $B$ components) repulsive walls with
$r^{-9}$ (unless stated otherwise) potential. Periodic boundary conditions were used in the $x$ and $y$ directions.
The interaction potential has the form
\begin{eqnarray}
\frac{V_{AB}(r)}{4 \epsilon_{AB}}&=&
\left \{
\begin{array}{ll}
 \Big[ \left(\dfrac{\sigma_{AB}}{r}\right)^{12} -  \left(\dfrac{\sigma_{AB}}{r}\right)^6   \Big], &  r <2.5\sigma_{AB} \\
 0,                                                                     &  r \geq 2.5\sigma_{AB} 
\end{array}
\right .
\label{KALJ}
\end{eqnarray}
The length and energy parameters used in the present study are 
$\epsilon_{AA}=1.0,\epsilon_{BB}=0.5,\epsilon_{AB}=0.25$ and 
$\sigma_{AA}=1.0,\sigma_{BB}=0.88,\sigma_{AB}=0.80$ unless stated
otherwise. All these parameters except $\epsilon_{AB}$ are the same as those in the Kob-Andersen 
Lennard-Jones (KALJ) model\cite{Kob1994PRL} in which $\epsilon_{AB}=1.50$.
Masses of all particles were taken to be the same ($M$).
All quantities are expressed in reduced units: the length, 
time, and temperature are expressed in units of $\sigma_{AA}$, 
$(M\sigma_{AA}^2/\epsilon_{AA})^{1/2}$, and $\epsilon_{AA}/k_B$.
In terms of $\epsilon_{AA}$, $\epsilon_{AB}$, and $\epsilon_{BB}$,
the mixing energy can be expressed as $\omega=[(\epsilon_{AA} + \epsilon_{BB})/2 - \epsilon_{AB}]$.
Thus, the value of mixing energy of the BLJ mixture is $\omega=+0.50$, and that of the KALJ mixture is
$\omega=-0.75$.

We have used a MD time step $dt=0.005$ and equilibration and data production runs
are $1$ and $4 \times 10^{6}$ MD steps long, respectively. We also performed canonical MC simulations for the BLJ system 
confined in the $z$ direction by two hard walls (particles are not allowed within a distance of
$0.50\sigma_{AA}$ from the walls). These simulations were carried out for a system 
with 1200 particles and an average step size of 0.14. 
{The BLJ potential was truncated at $2.50\sigma_{\alpha\beta} ~ [(\alpha,\beta)\in (A,B)]$ and shifted to ensure the continuity of the potential at the cutoff in the MC simulations. 
For surface tension calculations of single-component
Lennard-Jones (LJ) systems, we have used 4000 particles and a potential cutoff of $8.5\sigma$, 
where $\sigma$ is the particle diameter.
\section{Simulation Results}
In this section, we present the detailed results of our simulations of the equilibrium behavior of different kinds
of binary liquids near interfaces generated by repulsive walls with different potentials.

\subsection{Violation of the Guggenheim relation for positive mixing energy}

\begin{figure}[t]
\includegraphics[scale=0.45,angle=0]{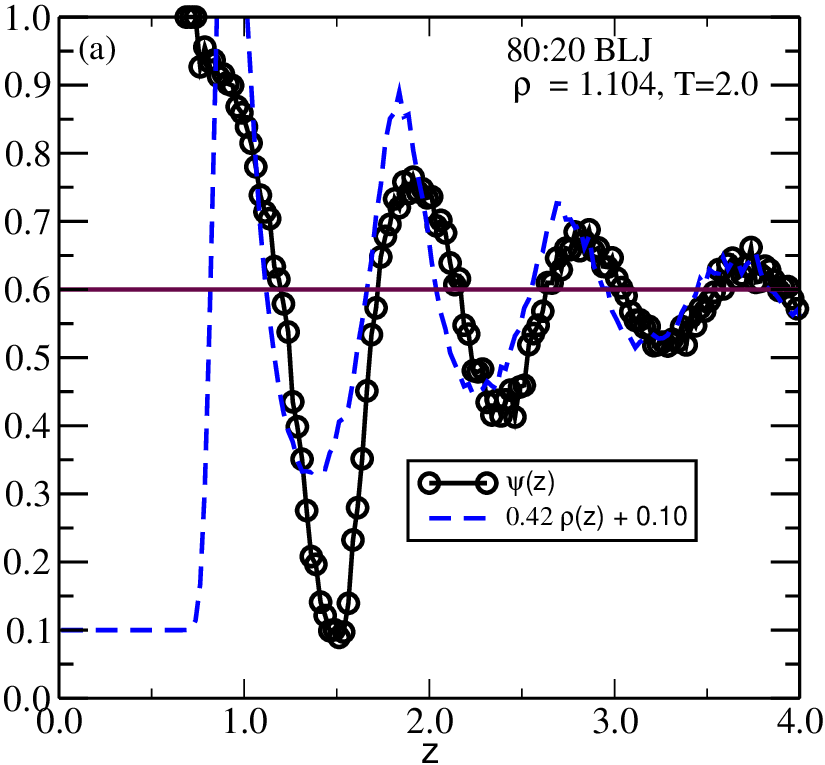}\vspace{10mm}
\includegraphics[scale=0.45,angle=0]{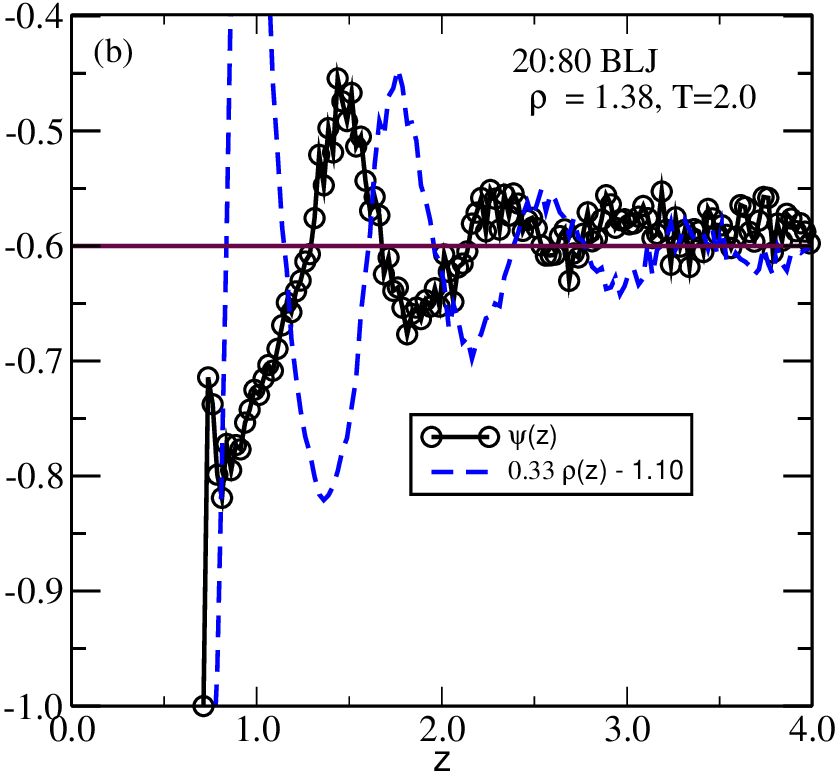}
\caption{(Color online) Order parameter $\psi(z)$ of the BLJ mixture confined 
by two repulsive $r^{-9}$ walls for (a) 80:20 and (b) 20:80 compositions. 
The temperature is $T=2.00$ and densities are (a) $\rho=1.104$  and
(b) $\rho=1.38$. The scaled and shifted density profile is also shown.}
\label{md}
\end{figure}

We first present the results for the BLJ mixture that has positive mixing energy.
Figure \ref{md}(a) exhibits the variation of the order parameter defined as 
$\psi(z)=[\rho_A(z)-\rho_B(z)]/[\rho_A(z)+\rho_B(z)]$ in the confinement ($z$) direction 
for a bulk 80:20 composition. The order parameter
can be rewritten as $\psi(z)=\rho_A(z)/[\rho_A(z)+\rho_B(z)] - \rho_B(z)/[\rho_A(z)+\rho_B(z)]$,
which is the difference between the fractional compositions of $A$ and $B$. For a 80:20
composition, the value of $\psi(z)$ will be 0.60 far away from the walls, where the system 
shows bulk behavior. Values of $\psi(z) > 0.60$ and $\psi(z)<0.60$ indicate
enrichment of the $A$ and $B$ components, respectively. 
The order parameter for the 20:80 composition is depicted in Fig. \ref{md}(b). The 
temperature is $T=2.0$ and bulk densities are 
$\rho=1.104$ in Fig. \ref{md}(a) and $\rho=1.38$ in Fig. \ref{md}(b). Different densities were considered in the two cases
in order to maintain the same volume fractions. It should be noted that the saturation density for large
values of $z$ is slightly higher than the bulk density $\rho$. This happens because of the formation of 
regions with no particles (vacuum) near the two
repulsive walls  (details can be found in Ref.\cite{Shibu-preprint1}). The 
$z$ dependence of the total density, scaled and shifted by arbitrary numbers 
[0.42 and 0.10 in Fig.~\ref{md}(a) and $0.33$ and $-1.10$ in Fig.~\ref{md}(b)], are also 
 shown in these figures to illustrate the familiar spatial modulation of the total density.

In Fig. \ref{md}(a) the $A$ component enriches the
interface while the $B$ component enriches the interface in Fig. \ref{md}(b); $A$ and $B$ 
are the majority components in Figs.~\ref{md}(a) and \ref{md}(b), respectively. Thus,
regardless of the values of the surface tension, the majority component 
enriches the interface for this BLJ system for which the
mixing energy $\omega=+0.50$, is positive. According to the Guggenheim adsorption relation, the minority
component should have enriched the interface for this system.
This observation is thus  in contradiction of the Guggenheim relation. 

\begin{figure}[b]
\includegraphics[scale=0.45,angle=0]{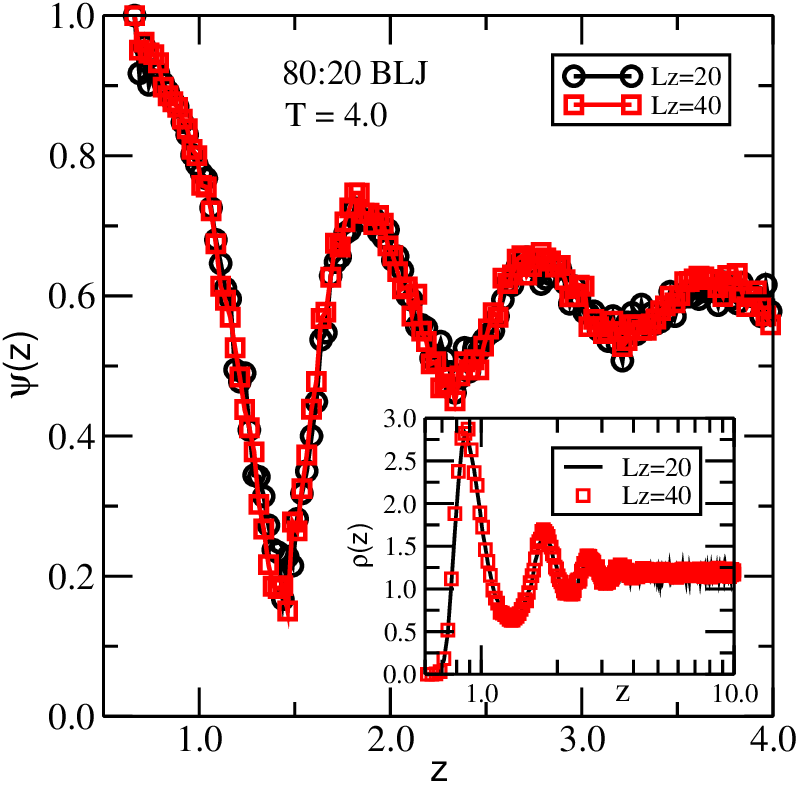}
\caption{(Color online) Comparison of the order parameter profile $\psi(z)$ of the 80:20 BLJ mixture confined 
by two repulsive $r^{-9}$ walls for $L_z=20$ with $N=2400$ and $L_z=40$ with $N=4800$ 
at temperature $T=4.0$. The inset shows a comparison of the $z$ dependence of the density $\rho(z)$ for 
$L_z=20$ and $40$. The density of a simulation cell with boundaries at $z=0.80$ and 
$z=L_z-0.80$ is $1.20$.} 
\label{md_diffN}
\end{figure}

To check whether the behavior described above is affected by the distance between the
walls (the width of the slit pore), we have simulated the behavior of the 80:20 BLJ system at $T=4.0$ for two 
values, 20 and 40, of $L_z$, the distance between two walls. 
In this comparison, it is necessary to take into account a trivial effect
of system size that was mentioned above. 
Since the density is vanishingly small in a region of width $\sim 0.8$
near each of the two repulsive walls, the constant saturation density in the middle of the
simulation cell is slightly higher than the bulk density and depends weakly on $L_z$ because
the vacuum regions near the walls correspond to a larger fraction of the total volume for smaller
values of $L_z$. To eliminate this trivial effect, we have compared the density profiles for
$L_z=20$ and $40$ obtained from simulations in which the repulsive walls were placed at distances
of $0.8$ away from the boundaries of the simulation box (the left and right boundaries of
the simulation box are at $z=0.80$ and $z=L_z-0.80$, respectively, whereas the repulsive walls
are at $z=0$ and $z=L_z$). The number of particles is  $N=2400$ for $Lz=20.0$ 
and $N=4800$ for $Lz=40.0$. Figure \ref{md_diffN} shows the comparison of $\psi(z)$ for these two 
system sizes. There is no significant difference in the two sets of data.
In the inset we show the $z$ dependence of the density $\rho(z)$ for the two 
values of $L_z$. No dependence on $L_z$  is seen and therefore $L_z=20$ is large enough to 
neglect the effect of the second wall on the modulation of the density  and the order parameter near the first 
one. 

The violation of the Guggenheim relation described above 
occurs for all compositions between 20:80 and 80:20, as 
shown in Fig. \ref{xA-Gug-MD}(b). Figure \ref{xA-Gug-MD} depicts the surface concentration 
$x'$ { vs} the bulk concentration $x$ of component $A$
for negative [Fig.~\ref{xA-Gug-MD}(a)] and  positive [Fig.~\ref{xA-Gug-MD}(b)] mixing energies. 
The values of $x'$ in MD simulations have been obtained from the concentration 
of $A$ at the position of the first peak in the density modulation.  The error bars reflect variations of the results obtained
from five independent MD simulations.
The bulk density for the 80:20 BLJ system 
is $\rho=1.104$ and for other compositions it has been adjusted to maintain 
the same volume fraction. The results shown in  
Fig. \ref{xA-Gug-MD}(a) are for the KALJ 
system \cite{Kob1994PRL} with negative mixing energy, $\omega=-0.75$. The simulation results are 
compared with the predictions of the Guggenheim relation in both cases. We emphasize the qualitative behavior (enrichment of
the surface layer by the majority or the minority component) in this comparison.
The Guggenheim relation qualitatively holds well for the KALJ system with negative mixing 
energy at low temperature $T = 1.0$ and at the high temperature $T = 4.0$: Both the Guggenheim 
relation and MD simulations exhibit enrichment of the surface layer by the majority component for all compositions. 
However, this relation is qualitatively violated for all compositions for the BLJ system with positive mixing energy at temperature $T = 4.0$.  
Figure \ref{xA-Gug-MD}(b) shows that while the minority component is supposed to enrich the interface 
according to the Guggenheim relation, the majority component is found to enrich the interface in MD simulations. 

The surface tensions of pure $A$ and pure $B$ LJ liquids are required for determining the values of $x'$ predicted by the
Guggenheim relation. The surface tension has been obtained from the difference of normal and
tangential pressure tensors~\cite{ST-Frenkel,ST-Sinha,ST-Vega}
\begin{eqnarray}
\gamma &=& \int_{phase1}^{phase2} [P_N(z) - P_T(z)] dz \\
&=& \left<  \sum_{i > j} \sum_{j} \dfrac{x_{ij}^2 + y_{ij}^2 -2 z_{ij}^2}{2\mathcal{A}r_{ij}} V_{ij}' -\sum_{j} \dfrac{z_j v'(z_j)}{\mathcal{A}}   \right>,
\end{eqnarray}
where $v(z)$ is the wall potential at distance $z$ from the wall and $\mathcal{A}$ is the
area of the simulation box normal to the $z$ direction. Unlike the
liquid-vapor interface, the shape of the wall-induced interface is not of the hyperbolic 
tangent form due to the presence of the wall. Therefore, it is difficult to evaluate the 
tail correction for the surface tension. We have used a large cutoff  of $8.5\sigma$ 
in this calculation, $\sigma$ being the particle diameter of
the one-component LJ system. Since the surface tension of the 
one-component system does not scale with the energy and size parameters in the 
presence of walls, we have calculated the surface tensions of pure $A$ and 
$B$ liquids independently. We have used $18.28 \sigma_{AA}\times 18.28\sigma_{AA} \times20.00\sigma_{AA}$ 
and $26.92 \sigma_{AA}\times 26.92\sigma_{AA} \times20.00\sigma_{AA}$
simulation cells with 4000 particles to calculate the surface tensions 
at temperatures $T = 1.0$ and $T = 4.0$, respectively. 
For these system sizes, finite-size capillary effects are negligible in the calculation of the surface tensions.
As the surface tension 
of a wall-induced interface strongly depends on the density profile, we
have used systems with significant density modulation at each temperature. 
Specifically, single-component LJ densities of 0.598 
and 0.276 were used for the evaluation of surface tensions at $T = 1.0$ and $4.0$, respectively. 
The diameter of the cross-sectional area parameter $a$ in Eq. \eqref{G2} was set to unity in the calculation of $x'$. 
The error bars shown in Fig.~\ref{xA-Gug-MD} for the data points representing the results obtained from
the Guggenheim relation reflect uncertainties in the calculated values of the surface tension.

\subsection{Interfaces near a structured wall}

So far, we have discussed the properties of wall-induced interfaces near structureless repulsive walls. 
We have found that the Guggenheim relation is qualitatively violated in mixtures with positive $w$ for interfaces created near 
structured walls as well. 
Figure \ref{md-fcc} exhibits the $z$ dependence of the order parameter for 80:20 
and 20:80 compositions of BLJ systems confined by two fcc walls. The fcc walls
consist of $A$ particles that
interact with $A$ and $B$ particles in the liquid in the same way.
All wall particles are fixed in a fcc structure and the
strength of the interaction between a particle in the wall and a fluid particle is 0.7 times 
that between two $A$ particles in the fluid. The particle densities in the walls for ithe 80:20 and 20:80 BLJ systems are 2.57 and 
3.62, respectively. The majority component is found to enrich the interface in both 80:20
and 20:80 mixtures, which is in disagreement with the prediction of the Guggenheim relation.

\begin{figure}[t]
\includegraphics[scale=0.45,angle=0]{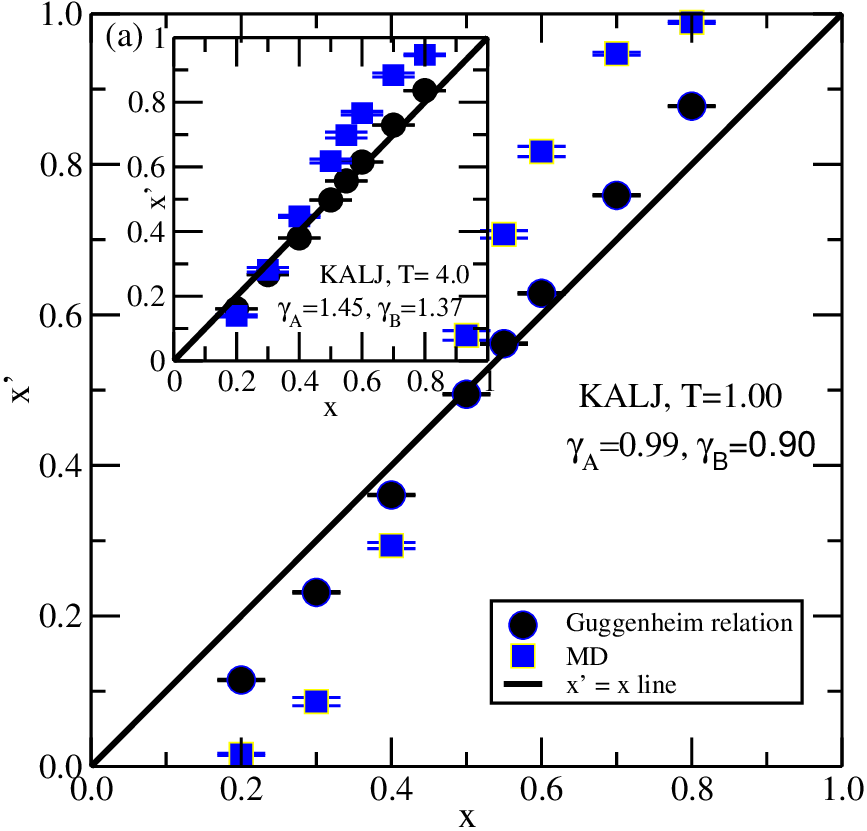}\vspace{10mm}
\includegraphics[scale=0.45,angle=0]{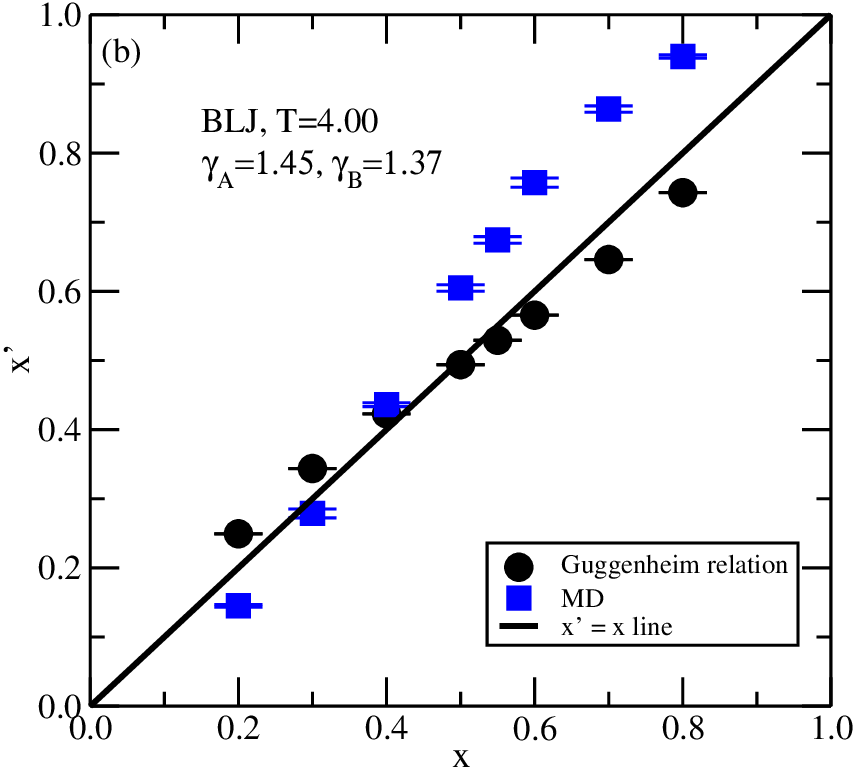}
\caption{(Color online) Plots of the interface concentration $x'$ { vs} the bulk concentration $x$ of 
the $A$ component for BLJ systems having mixing energies (a) $\omega=-0.75$ and 
(b) $\omega=+0.50$. Results of MD simulations are compared with the prediction of the Guggenheim relation. 
The surface tensions of $A$ and $B$ are 
0.99 and 0.90 for $T = 1.0$ and 1.45 and 1.37 for $T = 4.0$. The inset shows a  comparison at 
$T = 4.0$ for the KALJ system.}
\label{xA-Gug-MD}
\end{figure}

\begin{figure}[t]
\includegraphics[scale=0.45,angle=0]{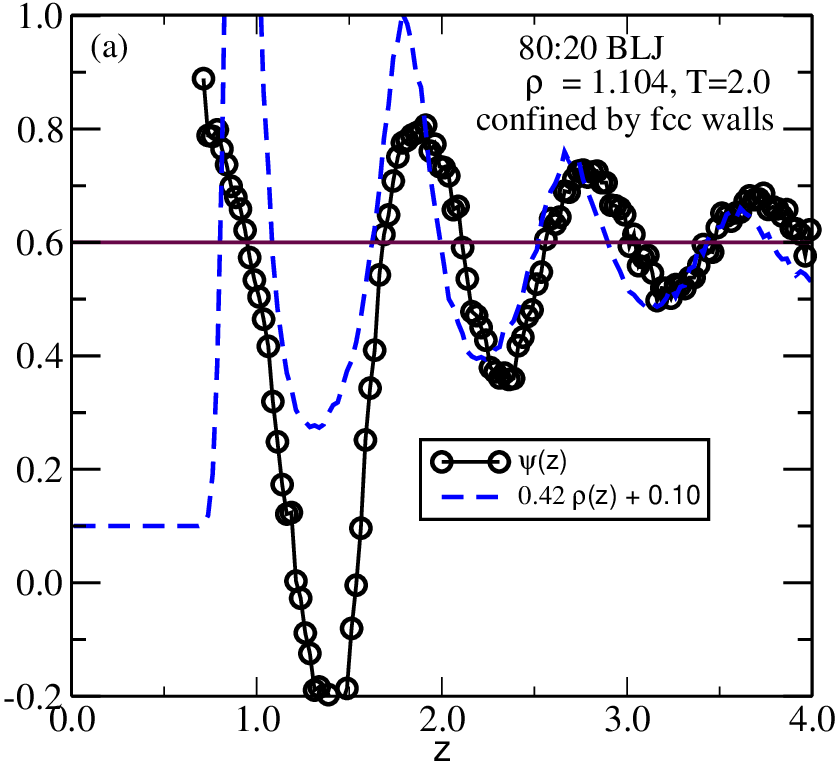}\vspace{10mm}
\includegraphics[scale=0.45,angle=0]{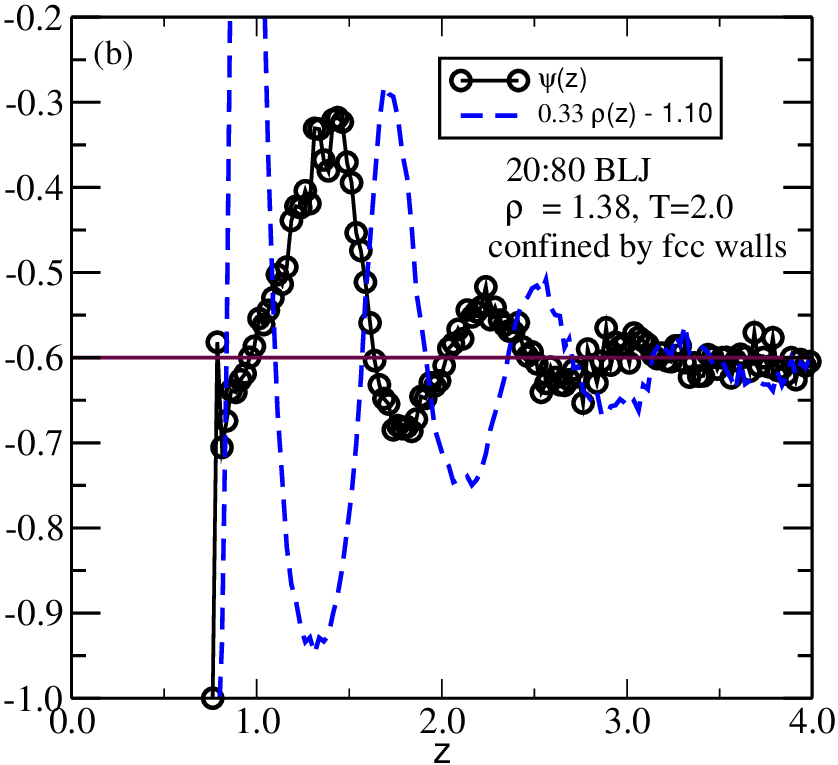}
\caption{(Color online) Order parameter $\psi(z)$ of a BLJ mixture confined 
by two fcc walls for (a) 80:20 and (b) 20:80 compositions. 
The temperature is $T=2.00$ and densities are (a) $\rho=1.104$  and
(b) $\rho=1.38$. The scaled and shifted density profile 
is also shown. The walls are of fcc 
structure, consisting of particles that interact with $A$ and $B$ 
particles of the fluid in the same way.}
\label{md-fcc}
\end{figure}

\subsection{Interfaces in binary mixtures of symmetric components}

We have considered another binary LJ mixture in which the constituent components have the same
size and energy parameters, {i.e.}, $\sigma_{AA}=\sigma_{AB}=\sigma_{BB}=1.0$
and $\epsilon_{AA}=\epsilon_{BB}=2\epsilon_{AB}=1.0$. Thus the mixing 
energy is $\omega=+0.50$, the same as that in the asymmetric BLJ mixture considered above. Since the size and
energy parameters of the two components are identical, the surface tensions of the components are the same and therefore the Gibbs 
adsorption rule fails to predict which component will enrich the interface for such 
a system. However, the Guggenheim relation predicts the enrichment of the minority component 
at the interface as the mixing energy is positive. At low densities and temperatures ({e.g.}, for 
$\rho=0.32$ and $T=0.95$, results not shown), our simulations of a system without confining walls 
show the formation of a liquid-vapor interface that is enriched by the minority component, in agreement with 
the prediction of the Guggenheim relation. If the system is confined by repulsive walls, 
enrichment of the minority component is seen at the wall-induced interface
[Fig.~\ref{puri}(a)] at density $\rho=0.92$ and temperature $T=1.50$.
If the density is increased at the same temperature, the system separates into 
$A$-rich and $B$-rich phases. If, however, the temperature is also increased to avoid
phase separation, the wall-induced interface is enriched by the majority 
component in violation of the Guggenheim relation, as shown in Fig.~\ref{puri}(b) 
for $\rho=1.38$ and $T=10.0$. 
If the temperature is increased at a fixed density, $\rho=0.92$, 
the value of the order parameter $\psi(z)$ at the interface increases from $\psi(z)=0.30$ at $T=1.50$ 
to approach $\psi(z)=0.60$ at $T=10.0$ (data not shown here). If the density 
is increased at $T=10.0$, the value of $\psi(z)$ starts deviating from $\psi(z)=0.60$ 
to higher values near a threshold density $\rho=1.104$ (data not shown here). 
So, at low densities and temperatures, the minority 
component enriches the liquid-vapor or wall-liquid interface in accordance
with the Guggenheim relation.  However, the majority component enriches the wall-liquid 
interface, in violation of the Guggenheim relation, at large densities and temperatures that
are sufficiently high to prevent phase separation in the bulk.
These findings show that the asymmetry in the energy and size parameters of the components of
the BLJ mixture is not responsible for the observed violation of the Guggenheim relation.

\begin{figure}[t]
\includegraphics[scale=0.38,angle=0]{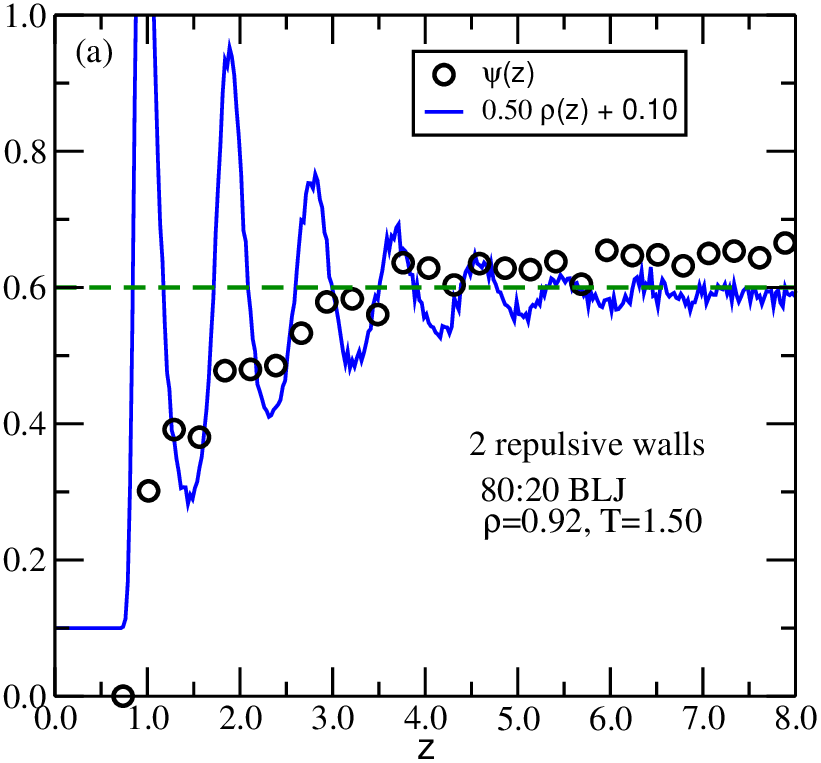}\vspace{8mm}
\includegraphics[scale=0.38,angle=0]{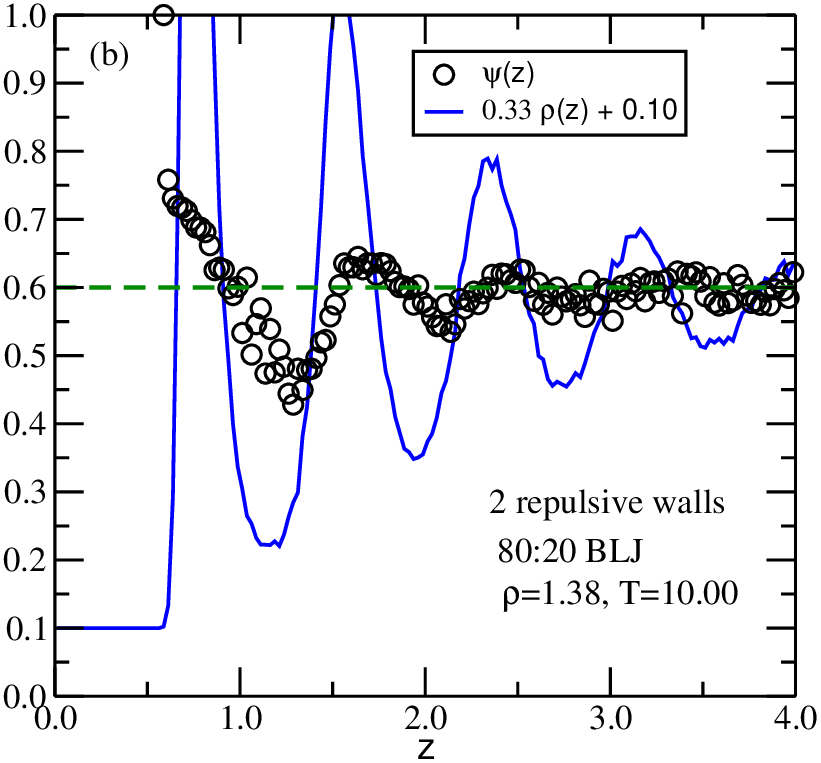}
\caption{(Color online) Order parameter $\psi(z)$ of a symmetric 80:20 binary LJ mixture with positive mixing energy,
confined by two $r^{-9}$ walls at (a) $\rho=0.92$ and $T=1.50$ and (b) $\rho=1.38$ and $T=10.0$ . Scaled and shifted profiles 
of the total density are also shown.
The length and energy parameters are $\sigma_{AA}=\sigma_{BB}=\sigma_{AB} $=1.00 and 
$\epsilon_{AA}=\epsilon_{BB}=2\epsilon_{AB} $=1.00, respectively. }
\label{puri}
\end{figure}
\subsection{Effects of density on the composition near wall-induced interfaces}
\begin{figure}[t]
\includegraphics[scale=0.38,angle=0]{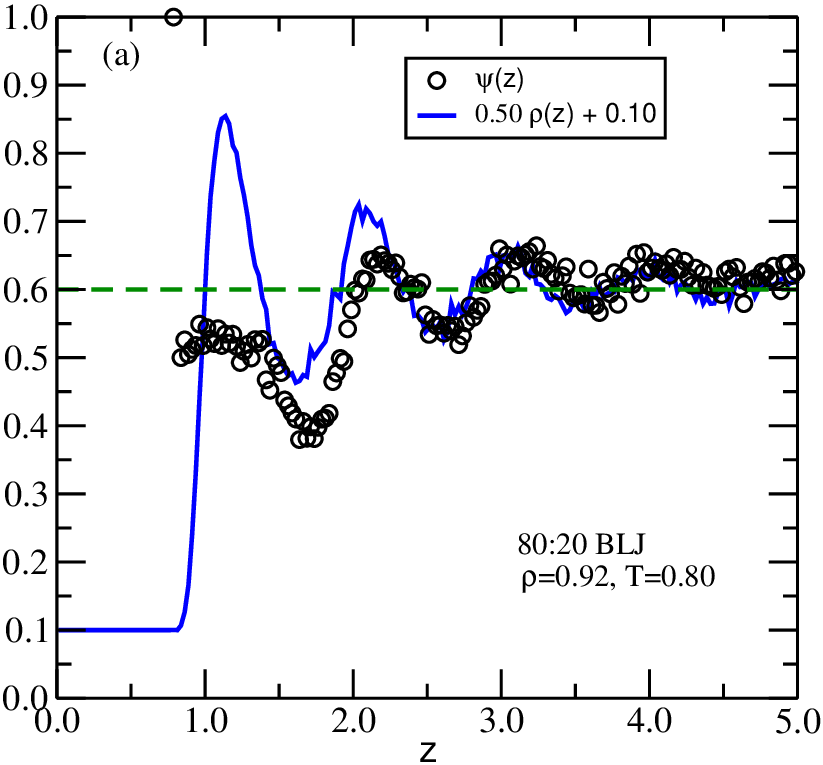}\vspace{8mm}
\includegraphics[scale=0.38,angle=0]{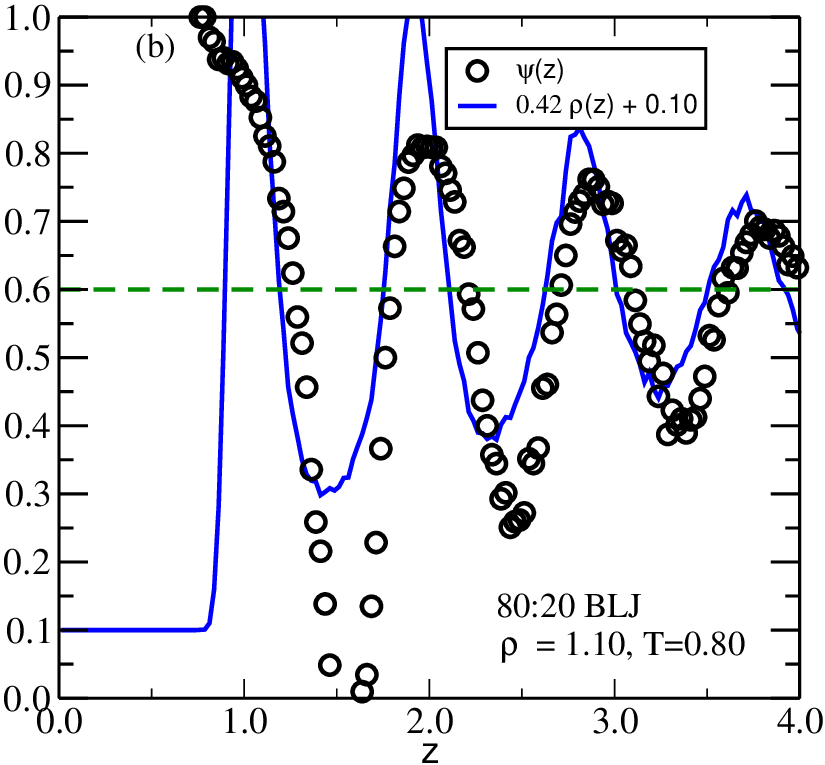}
\caption{(Color online) Order parameter $\psi(z)$ of an 80:20 BLJ mixture confined 
by two $r^{-9}$ walls at temperature $T = 0.80$ and densities (a) $\rho=0.92$ and (b) $\rho=1.10$. Scaled and shifted profiles 
of the total density are also shown. }
\label{rho0.8}
\end{figure}

The results described in the preceding section suggest that the Guggenheim relation is violated in
systems with positive mixing energy if the density is large. To check the generality of this observation, we
have examined the behavior of the 80:20 BLJ mixture confined by two $r^{-9}$ repulsive walls at different densities,
keeping the temperature fixed.
Figure \ref{rho0.8} exhibits the $z$ dependence of the order parameter $\psi(z)$ 
at $T=0.80$ and bulk densities  $\rho=0.92$ [Fig.~\ref{rho0.8}(a)] and  $\rho=1.10$ [Fig.~\ref{rho0.8}(b)]. The mixing energy is positive in this system and 
therefore it exhibits a tendency to separate into $A$-rich and $B$-rich phases. The temperature and densities considered
in these simulations were chosen to avoid bulk phase separation. At the lower density ($\rho=0.92$), 
the minority $B$ component enriches the wall-induced
interface, in accordance with the prediction of the Guggenheim relation.
It should be noted that $B$ has lower surface tension compared to $A$ and therefore 
the surface tension plays an important role in the enrichment of the surface layer by $B$.
However, it is evident from Fig. \ref{rho0.8}(b) that as the
density is increased, the dominance of the minority component at the wall-induced interface disappears and
at density $\rho=1.10$, the
majority component enriches the interface, violating the prediction of the Guggenheim
relation. Thus, the Guggenheim relation remains valid at low densities, but
it is violated at high densities. 
\subsection{MC simulation results for confinement by hard walls}
In order to examine if the nature of the interaction of the fluid particles with the walls plays any role in the violation 
of the Guggenheim relation, we replace the repulsive $r^{-9}$ walls by hard 
walls, which do not allow a particle to come closer than $0.5\sigma_{AA}$.
As MD simulations are difficult for this situation, we perform MC simulations
at constant temperature and density.  
The $z$ dependence of the order parameter $\psi(z)$ and the scaled and shifted 
total density $\rho(z)$ obtained from our MC simulations are shown for 80:20 and 
20:80 BLJ mixtures in Figs. \ref{mc}(a) and \ref{mc}(b) at temperature $T=4.0$ 
and densities $1.10$ and $1.38$, respectively. The confining walls here are hard walls 
that do not interact with the particles in the BLJ system as long as they do not violate the
hard-wall constraint. In this case also, the majority component dominates the interface for the
BLJ mixture with positive mixing energy,
violating the prediction of the Guggenheim relation. Thus the nature of the interaction with the walls does not play any
significant role in the violation of the Guggenheim relation. \vspace{8mm}
\begin{figure}[t]
\includegraphics[scale=0.45,angle=0]{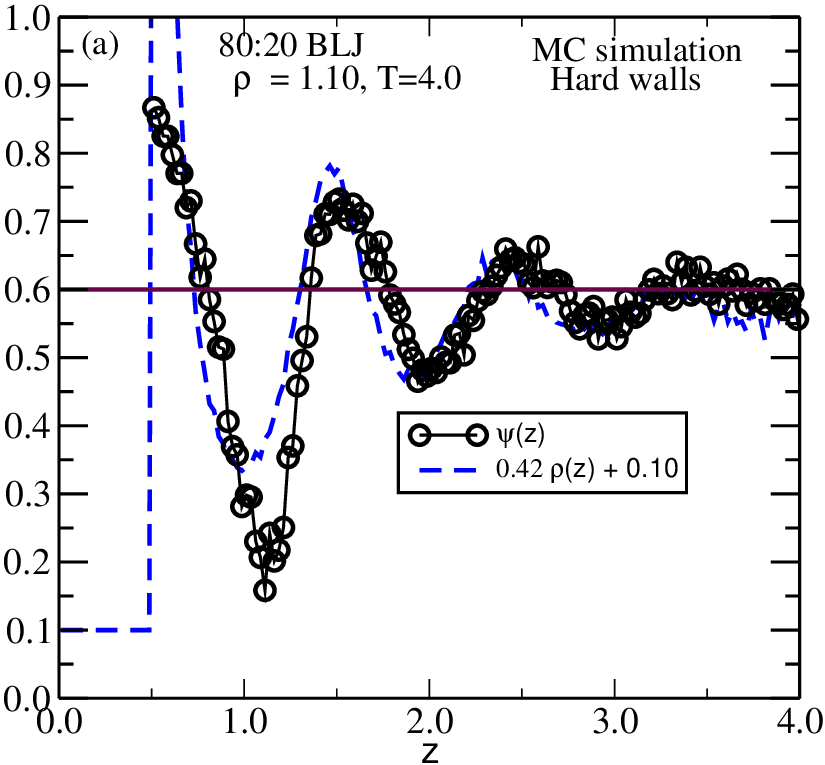}\vspace{10mm}
\includegraphics[scale=0.45,angle=0]{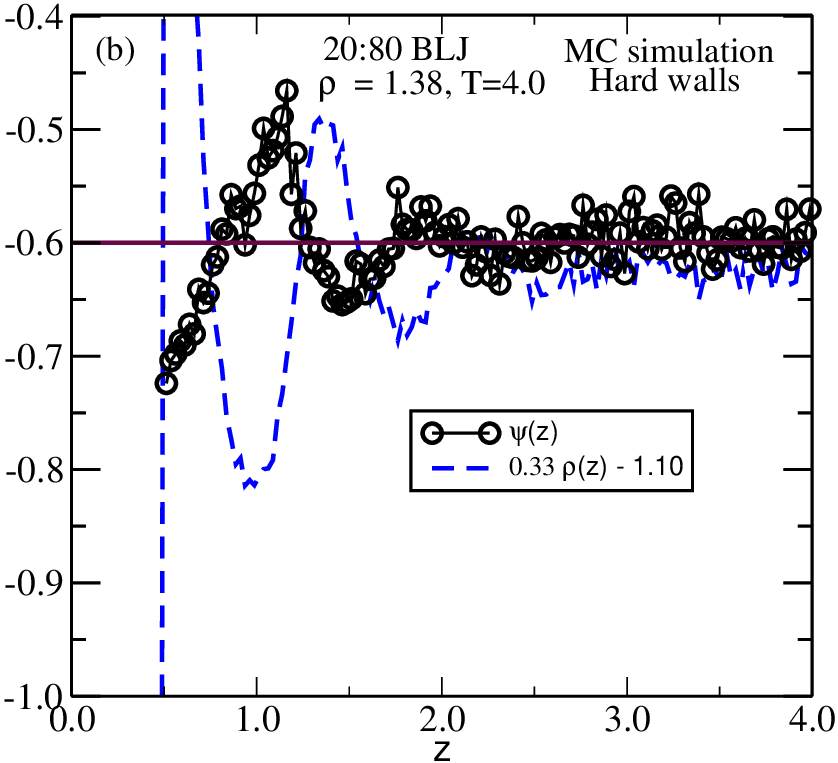}
\caption{(Color online) Order parameter $\psi(z)$ obtained from MC simulations for 
(a) 80:20 and (b) 20:80 BLJ mixtures confined by hard walls at temperature $T=4.0$ 
and densities (a) $\rho=1.10$  and (b) $\rho=1.38$. Scaled and shifted profiles of the total density are also shown.}
\label{mc}
\end{figure}
\section{Results Obtained from Density Functional Theory}

Classical DFT has been very successful~\cite{dftrev}  in providing 
a theoretical understanding of equilibrium properties of inhomogeneous
liquids. In order to explore the reason(s) for the observed violation of the Guggenheim relation for wall-induced interfaces in 
binary liquids with positive mixing energy, we have used DFT to calculate the density profiles of the two components 
of a binary liquid near a repulsive wall. In DFT, the
grand potential functional 
$\Omega[\rho({\bf r})]$ of a single-component liquid in the presence of an external potential $V({\bf r})$
is written as
\begin{equation}
\Omega[\rho({\bf r})] = F_{id}[\rho({\bf r})] + F_{ex}[\rho({\bf r})] + \int (V({\bf r})-\mu)\rho({\bf r})d{\bf r},
\label{grandpot1}
\end{equation}
where $F_{id}[\rho({\bf r})]$ and $F_{ex}[\rho({\bf r})]$ are, respectively, the ideal gas part and the excess part, arising from
interactions, of the Helmholtz free energy, and $\mu$ is the chemical potential. In the absence 
of any external potential, the equilibrium density is uniform and the grand 
potential function can be written as
\begin{equation}
\Omega[\rho] = F_{id}[\rho] + F_{ex}[\rho] - \int \mu\rho d{\bf r},
\label{grandpot2}
\end{equation}
where $\rho$ is the equilibrium bulk density. 
Equations \eqref{grandpot1} and \eqref{grandpot2} readily yield
\begin{equation}
\Omega[\rho({\bf r})]-\Omega[\rho] = \mathcal{F}^{ex} - \int \mu \delta\rho({\bf r})d{\bf r},
\end{equation}
where $\delta\rho({\bf r})=\rho({\bf r})-\rho$ and $\mathcal{F}^{ex}$ is given by
\begin{equation}
\mathcal{F}^{ex} = \Delta F_{id}[\rho] + \Delta F_{ex}[\rho] + \int V({\bf r})\rho({\bf r})d{\bf r},
\label{functional}
\end{equation}
with $\Delta F_{id}[\rho] \equiv F_{id}[\rho({\bf r})] - F_{id}[\rho]$ 
and $\Delta F_{ex}[\rho] \equiv F_{ex}[\rho({\bf r})] - F_{ex}[\rho]$.

We have used the simple Ramakrishnan-Yussouff (RY) \cite{RY} free-energy functional in our calculations 
for confined BLJ mixtures.
The RY functional for a single-component system without an external potential has the form
\begin{eqnarray}
\beta \mathcal{F}^{ex}[\rho({\bf r})] &=& \int d{\bf r} [ \rho({\bf r}) \ln\frac{\rho({\bf r})}{\rho} - \delta \rho({\bf r})] \nonumber \\
 &-& \frac{1}{2}\int \int d{\bf r} d{\bf r'} C(|{\bf r}-{\bf r'}|) \delta\rho({\bf r}) \delta \rho({\bf r'}), \label{ryeq}
\end{eqnarray} 
where $\beta=(k_BT)^{-1}$ and $C(r)$ is the direct pair correlation function of the uniform liquid with density $\rho$.
For the case of a binary mixture, in the presence of external 
potential due to a wall, the RY functional becomes
\begin{eqnarray}
&&\beta \mathcal{F}^{ex}[\rho_A({\bf r}),\rho_B({\bf r})] = \int d^3{\bf r} \Big[ \rho_A({\bf r})\ln\frac{\rho_A({\bf r})}{\rho_{A}} - \delta \rho_A({\bf r}) + \nonumber \\
&& \rho_B({\bf r})\ln\frac{\rho_B({\bf r})}{\rho_{B}} - \delta \rho_B({\bf r}) \Big] - \frac{1}{2}\int \int d^3{\bf r} d^3{\bf r'} \Big[C_{AA}(|{\bf r}-{\bf r'}|) \nonumber \\ 
&& \delta\rho_A({\bf r}) \delta \rho_A({\bf r'})+  2 C_{AB}(|{\bf r}-{\bf r'}|)\delta\rho_A({\bf r}) \delta \rho_B({\bf r'}) + \nonumber \\
&& C_{BB}(|{\bf r}-{\bf r'}|) \delta\rho_B({\bf r})\delta \rho_B({\bf r'}) \Big] + \beta \int d^3{\bf r} V({\bf r})[\rho_A({\bf r}) + \nonumber \\ 
&& \rho_B({\bf r})],
\label{dft1}
\end{eqnarray}
where $\delta\rho_A({\bf r})=\rho_A({\bf r})-{\rho_A}$ and $\rho_{A}$ is the bulk density for the $A$ component.
Similarly, $\delta\rho_B({\bf r})=\rho_B({\bf r})-\rho_{B}$ for the $B$ component. 
One must not equate the bulk density with the saturation density far away
from the walls.
Due to the presence of vacuum-like regions near the repulsive walls, the saturation density is not the 
same as the bulk density in finite samples, the former being slightly higher
than the latter (see Ref. \cite{Shibu-preprint1}).  
The direct correlation function $C_{\alpha,\beta}(r) (\alpha,\beta \in A,B)$ is obtained from integral 
equation theory employing Zerah-Hansen~\cite{ZH} closure. 

The extremum principal of DFT leads to the Euler-Lagrange equations 
\begin{eqnarray}
\frac{\delta}{\delta \rho_A({\bf r})} \mathcal{F}^{ex} = \mu_A, ~\frac{\delta}{\delta \rho_B({\bf r})} \mathcal{F}^{ex} = \mu_B,
\label{dft2}
\end{eqnarray}
where $\mu_A$ and $\mu_B$ are the chemical potentials of the two components. These equations 
are solved iteratively to yield $\rho_A(r)$ and $\rho_B(r)$.
The density is assumed to be uniform in the  $xy$ plane parallel to the walls.

\begin{figure}[t]
\vspace{4mm}
\includegraphics[scale=0.45,angle=0]{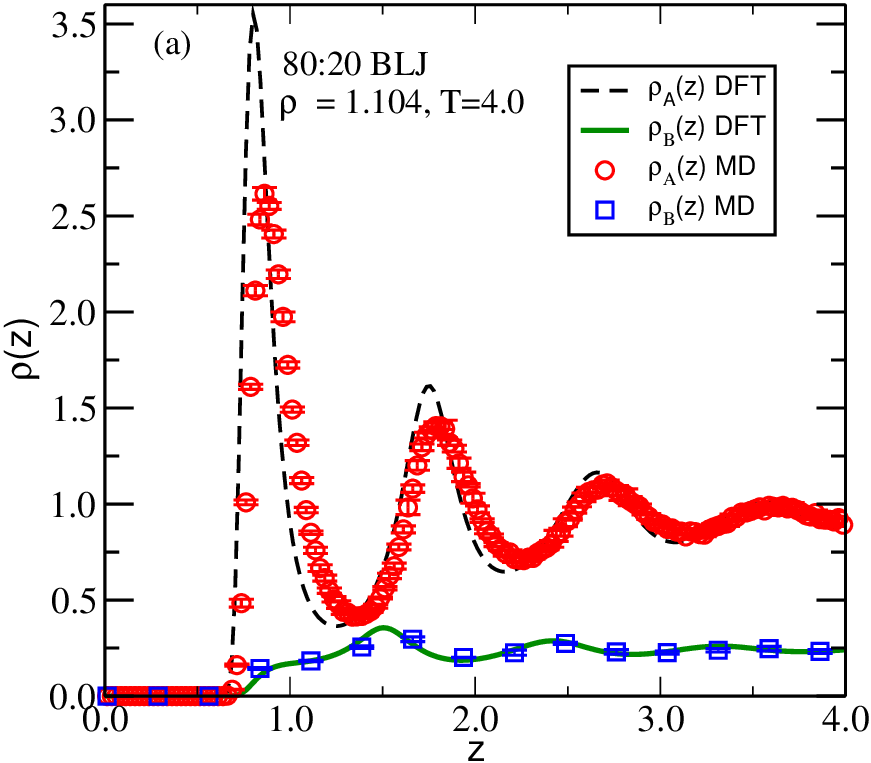}\vspace{10mm}
\includegraphics[scale=0.45,angle=0]{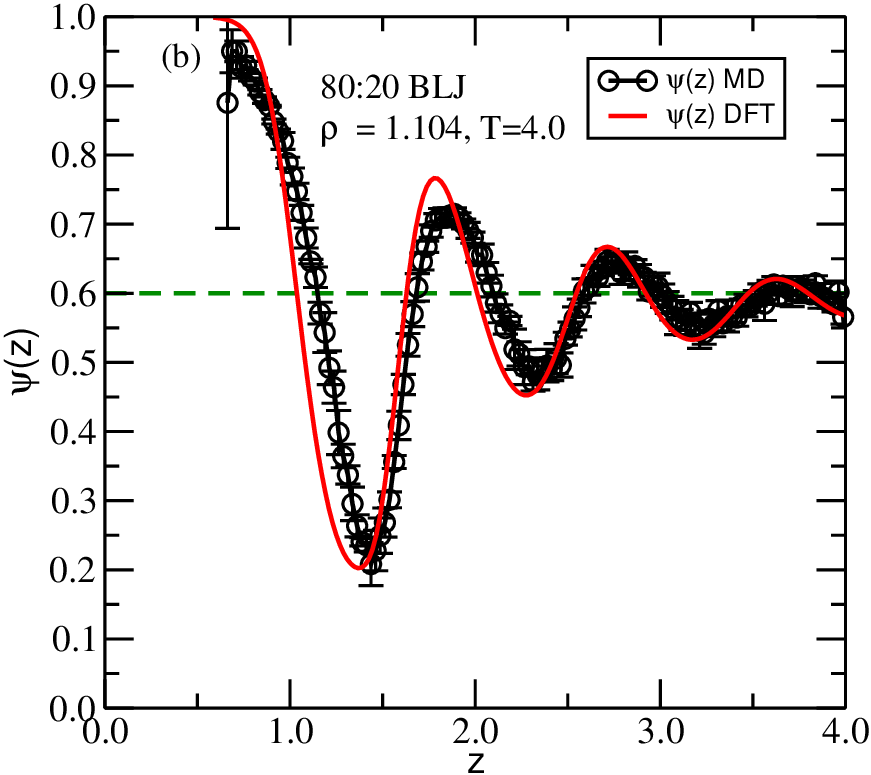}
\caption{(Color online) Comparison of the $z$ dependence of (a) the density profiles and (b) the order 
parameter $\psi(z)$ obtained from DFT calculations and MD simulations at 
density $\rho=1.104$ and temperature $T=4.0$ for an 80:20 BLJ system.}
\label{dft}
\end{figure}

Figure \ref{dft} shows a comparison of the density profiles of the two components and the order parameter 
obtained from DFT calculations and MD simulations for a BLJ system with an 80:20 
composition at density $\rho=1.104$ and $T=4.0$. It is clear that there is good agreement 
between the two sets of results. The discrepancy in the magnitude of the first density peak for the $A$ component  is probably
due to limitations of the RY functional for soft LJ potentials. 
Similar agreement between the results of DFT calculations and MD simulations is also found 
for the 20:80 BLJ system (data not shown here).
Note that the Guggenheim relation is violated in this BLJ system with positive mixing energy. 
Thus, the violation of the Guggenheim relation
found in the simulations is reproduced in the DFT calculations.
 
We have addressed the question of why the composition of the mixture near the walls is
different from that in the bulk. To separate out the contributions of the entropy and the internal energy, we write
the free-energy difference $\mathcal{F}^{ex}_{min}$ in terms of the entropy difference $S^{ex}$ and 
the difference in the internal energy $U^{ex}$:
\begin{equation}
\mathcal{F}^{ex}_{min} = U^{ex} - TS^{ex},
\label{free}
\end{equation}
where $\mathcal{F}^{ex}_{min}$ is the value of $\mathcal{F}^{ex}$ for
the density profiles obtained from the solution of Eq. (\ref{dft2}) and the excess entropy is given by 
\begin{eqnarray}
\label{entropy}
S^{ex} = -\frac{\partial}{\partial T} \mathcal{F}^{ex}_{min}.
\end{eqnarray}
This quantity is obtained from the slope of a plot of $\mathcal{F}^{ex}_{min}$ {vs} $T$.  

We have also carried out a calculation in which the composition of the mixture is 
forced to remain 
unchanged throughout the system, while the total density is allowed to vary with $z$
by applying an extra constraint $\rho_A({\bf r})/\rho_B({\bf r})= s$, 
where $s$ is the ratio of the densities of the two components in the bulk. 
With this constraint the RY functional can be rewritten as
\begin{eqnarray}
&&\beta \mathcal{F}^{ex}[s\rho_B({\bf r}),\rho_B({\bf r})] = \int d^3{\bf r} (s+1)\Big[ \rho_B({\bf r})\ln\frac{\rho_B({\bf r})}{\rho_{B}} \nonumber \\
&& - \delta \rho_B({\bf r})\Big] - \frac{1}{2}\int \int d^3{\bf r} d^3{\bf r'} \Big[s^2C_{AA}(|{\bf r}-{\bf r'}|)  \nonumber \\ 
&&  +  2 s C_{AB}(|{\bf r}-{\bf r'}|) + C_{BB}(|{\bf r}-{\bf r'}|) \Big]\delta\rho_B({\bf r})\delta \rho_B({\bf r'}) \nonumber \\ 
&& + \beta \int d^3{\bf r} V( {\bf r})(s +1) \rho_B({\bf r}).
\end{eqnarray}
For the sake of clarity of notation, we denote the Helmholtz free-energy 
functional for this case by $ \mathcal{F}^{ex}_c$. With this additional 
constraint, the excess grand potential $\Delta \Omega_c$ is given by
\begin{equation}
\Delta\Omega_c[\rho] =  \mathcal{F}^{ex}_c[s\rho_B({\bf r}),\rho_B({\bf r})] - \int (s\mu_A + \mu_B) \delta\rho_B({\bf r}))d^3{\bf r},
\end{equation}
and the extremum condition leads to the equation
\begin{eqnarray}
\label{cdft}
\frac{\delta}{\delta \rho_B({\bf r})} \mathcal{F}^{ex}_c = (s\mu_A +\mu_B) = \mu_c.
\end{eqnarray}

As expected, the optimal value of the Helmholtz free energy in the constrained case is 
higher than that in the unconstrained 
case, {i.e.}, $\mathcal{F}^{ex}_{c,min} > \mathcal{F}^{ex}_{min}$.
The total density profile $\rho({\bf r}) = [\rho_A({\bf r}) + \rho_B({\bf r})]$ 
obtained in the constrained minimization is slightly different from that obtained from the 
unconstrained minimization.
We have also calculated the Helmholtz free energy for a situation where the 
total density profile $\rho({\bf r})$ is exactly the same as that in the unconstrained 
case, but the local densities $\rho_A({\bf r})$ and $\rho_B({\bf r})$ are 
rescaled to maintain the ratio $s$ of the bulk densities. This is 
done as follows. From the solution of \eqref{dft2} we obtain $\rho_A({\bf r})$ 
and $\rho_B({\bf r})$ at a given temperature and density. We then replace 
$\rho_A({\bf r})$ by $s\rho({\bf r})/(s+1) $ and $\rho_B({\bf r})$ by $\rho({\bf r})/(s+1)$. 
The Helmholtz free energy in this case is denoted 
by $\mathcal{F'}^{ex}$. The values of the Helmholtz free energy for these three 
cases are found to be in the  order 
$\mathcal{F'}^{ex} > \mathcal{F}^{ex}_{c,min} > \mathcal{F}^{ex}_{min}$. 
Using Eqs. \eqref{free} and \eqref{entropy}, we estimate the 
contributions of the entropy and the internal energy for all the three cases. 

Figure~\ref{dft-free} exhibits the $T$ dependence of 
$\beta \Delta F(T) = \beta(\mathcal{F}^{ex}_{min} - \mathcal{F'}^{ex})$, 
$-\Delta S(T) = -(S^{ex} - S'^{ex})$, and $\beta \Delta U(T) = \beta (U^{ex} -U'^{ex})$  
for mixing energy $\omega = + 0.50$ at $T=4.0$ and $\omega=-0.75$ (inset) 
at $T=3.0$ at density $\rho=1.3872$ for the 20:80 composition, {i.e.}, $s=0.25$.
We prefer to compare $\mathcal{F}^{ex}_{min}$ with $\mathcal{F'}^{ex}$ 
because the total density profiles are exactly the same in these two cases and 
therefore, the contributions of the wall potential are also the same. We emphasize that the 
difference between $\mathcal{F'}^{ex}$  and $\mathcal{F}^{ex}_{c,min}$ is very small.  
For positive as well as negative mixing energy, the free energy in
the unconstrained case is, as expected, lower than that in the constrained situation where a 20:80 
composition is maintained throughout the system. From Fig. \ref{dft-free} 
it is also evident that the decrease in the free energy is due to a decrease in the internal energy. 
The entropy in the constrained case is higher than that in the unconstrained case because the two
components are less well mixed (the local composition is more heterogeneous) in the unconstrained case.
While the entropy favors the constrained situation with uniform composition, the internal energy 
favors the unconstrained one. The contribution of the internal energy towards decreasing the 
free energy is larger than that of the entropy towards increasing it and 
hence the free energy is lowered in the unconstrained case.

\begin{figure}[t]
\vspace{4mm}
\includegraphics[scale=0.45,angle=0]{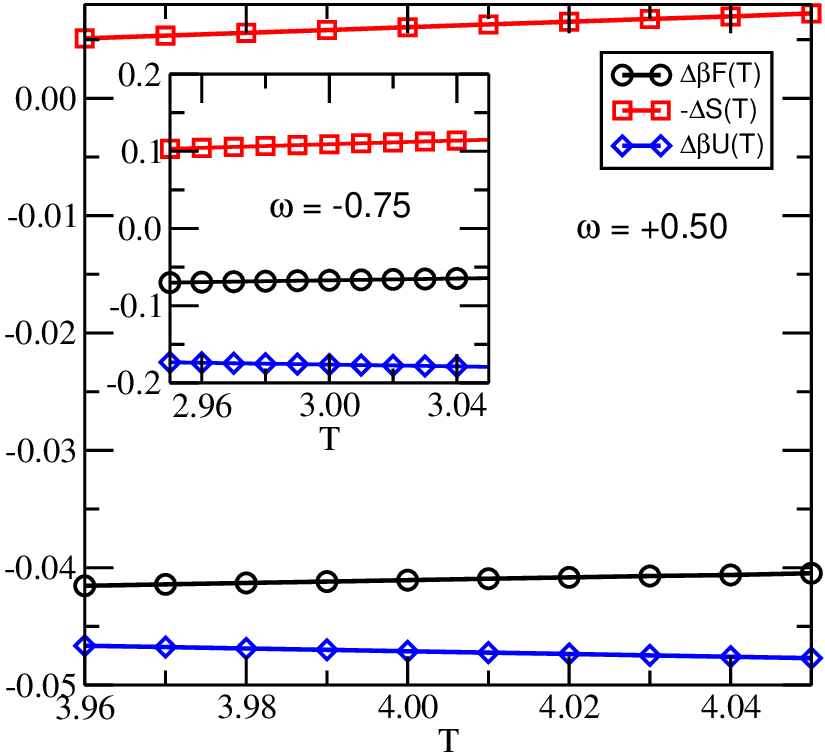}
\caption{(Color online) The $T$ dependence of $\beta \Delta F(T) = \beta(\mathcal{F}^{ex}_{min} - \mathcal{F'}^{ex}_{min})$ , $-\Delta S(T) = -(S^{ex} -S'^{ex})$, and $\beta \Delta U(T) = \beta (U^{ex} -U'^{ex})$ redistributed local densities in the ratio 20:80. The inset shows the same quantity but for negative mixing energy.}
\label{dft-free}
\end{figure}
\section{Summary and Discussion}

We have studied in detail the variation in the local composition of binary liquids near 
a repulsive wall that has the same potential for both components of the liquid. The observed behavior near the wall-induced
interface is compared with the prediction of the Guggenheim relation for vapor-liquid interfaces, which is a
theoretical description of surface-induced composition modulation.
When the mixing energy of the binary system is negative, {i.e.}, when the two components of the mixture tend to be mixed, 
the majority component is found to enrich the interface at all the
temperatures and densities considered in our simulations. This is in qualitative agreement with the prediction 
of the Guggenheim relation.
If the mixing energy is positive, phase separation occurs at low temperatures, but a homogeneous
bulk phase is present at higher temperatures. In this situation, the wall-induced interface is enriched by the
minority component, in agreement with the prediction of the Guggenheim relation, if the bulk density is low.
However, at high densities and temperatures high enough to prevent phase separation in the bulk, 
the majority component enriches the interface, in qualitative disagreement with the prediction of the 
Guggenheim relation. From simulations for different binary mixtures and wall potentials, we show that this
behavior is robust: It is found for both structured and structureless walls, for both symmetric and asymmetric mixtures
and for walls with both soft and hard repulsive potentials. A DFT calculation using the Ramakrishnan-Yussouff free-energy
functional reproduces the behavior found in our simulations, including the violation of the Guggenheim relation at 
high densities in binary liquids with positive mixing energy.

While we do not have a complete understanding of the reason(s) for the observed violation of the prediction of the Guggenheim relation, our
simulation and DFT results provide a few clues. As noted in Sec. II, the derivation of the Guggenheim relation involves several
approximations. The mean-field treatment used in the derivation of the Guggenheim relation does not appear to be the reason for its
violation because the DFT calculation, which is also a mean-field theory, yields results in agreement with those of simulations. The
main difference between the theory behind the Guggenheim relation and DFT is that the latter takes into account the effects of short-range
correlations in a dense liquid through the use of the direct pair correlation function in the free energy functional. The derivation of the 
Guggenheim relation outlined in Sec. II is based on the assumption that the total density is uniform and the composition is different
from that in the bulk only in one layer at the interface. Our simulations, on the other hand, show that the total density exhibits
considerable spatial modulation near the wall and the effects of the wall persist for several layers before bulk behavior is restored. Both
these effects arise from short-range correlations present in the liquid. These features are reproduced in the DFT calculation that takes
into account these correlations. However, these effects are completely neglected in the derivation of the Guggenheim relation. 
These observations suggest that the failure of the Guggenheim relation to predict the nature of the modulation of the 
composition near wall-induced interfaces in binary liquids with positive mixing energy is a consequence of  neglecting the effects of short-range
correlations present in the liquid. This conclusion is supported by the observation that the violation of the prediction of the Guggenheim relation
is found only for liquids with relatively high density for which short-range correlations are more pronounced; simulations for liquids
at the same temperature but at lower densities do not exhibit a qualitative violation of the  prediction of the Guggenheim relation.

As noted in the Introduction, variations in the composition of vapor-deposited glass films near the vacuum-glass and glass-substrate
interfaces are believed to play an important role in determining the physical properties of the glass. The process of vapor deposition has
been simulated in Ref.~\cite{dePablo2013condmat}, using the KALJ binary system. In that work, the majority component was found to enrich both
vacuum-glass and glass-substrate interfaces. This is consistent 
with our results and those of Ref. \cite{Osborne} since the mixing energy is negative in the 
KALJ system. In Ref.~\cite{dePablo2013condmat}, a spatial variation of the composition was also found in glasses obtained by
rapidly cooling the binary liquid on a substrate from a high temperature. These results indicate that a modulation of the composition near interfaces may be
a generic property of glasses with two or more components. A detailed investigation of how the properties of such glasses are affected
by the spatial variation of the composition near interfaces would be most interesting.  
\section*{ACKNOWLEDGEMENTS}
We thank S. P. Das, S. Puri, S. Sastry, and H. R. Krishnamurhty for fruitful
discussions. We acknowledge Department
of Science and Technology, India for financial support and the
Supercomputer Education and Research Center of Indian Institute of Science and Shiv Nadar University for computational
facilities.

\end{document}